\documentclass[twocolumn,showpacs,preprintnumbers,amsmath,amssymb]{revtex4}
\usepackage{dcolumn}
\usepackage{bm}

\usepackage{graphics}
\newcommand{\ds}{\displaystyle}

\newcommand{\be}{\begin{equation}}
\newcommand{\en}{\end{equation}}
\newcommand{\bea}{\begin{eqnarray}}
\newcommand{\ena}{\end{eqnarray}}
\begin{document}

\title{Closed inflationary universe with tachyonic field}
\author{Leonardo Balart}
\author{Sergio del Campo}
\author{Ram{\'o}n Herrera}
\author{Pedro Labra{\~n}a}
\affiliation{Instituto de F\'{\i}sica,
Pontificia Universidad Cat\'olica de Valparaiso \\
Av. Brazil 2950, Valpara\'{\i}so, Chile, Casilla 4950,
Valpara\'{\i}so.}

\begin{abstract}
In this article we study  closed inflationary universe
models by using a tachyonic field theory. We determine and
characterize the existence of an universe with $\Omega > 1$, and
which describes  a period of inflation. We find that considered
models are less restrictive compared to the
 standard ones  with a scalar field. We use recent
astronomical
 observations  to constraint  the parameters appearing in the model.
 Obtained results are compared to those found in the standard scalar field inflationary universes.

\end{abstract}


\maketitle

\section{Introduction}
\label{intro}
 The existence of Doppler peaks and their
respective localization tend to confirm the inflationary paradigm,
associated with a flat universe with $\Omega\simeq 1$, as
corroborated by the existence of an almost scale invariant power
spectrum, with $n_s\sim 1$ \cite{Peiris:2003ff,WMAP3}.

The recent temperature anisotropy power spectrum, measured with
the Wikinson Microwave Anisotropy Probe (WMAP three-year data) at
high multipoles, is in agreement with an inflationary $\Lambda$-
dominated CDM cosmological model. However, the low order
multipoles have lower amplitudes than expected from this
cosmological model \cite{Bennett:2003bz}, and the mismatch in
these amplitudes may indicate the need for new physics.
Speculations for explaining this discrepancy has been invoked in
the sense that the low quadrupole observed in the CMB is related
to the curvature scale \cite{Efstathiou:2003hk}. Also, the CMB
data \cite{MacT}, alone places a constraint on the curvature which
is $\Omega_k=-0.037_{-0.039}^{+0.033}$. Additions of the LSS data,
yields a median value of $\Omega_k=-0.027\pm 0.016$. Restricting
$H_0$ by the application of a Gaussian HST prior, the curvature
density determined from Boom2K flight data set and all previous
CMB results are $\Omega_k=-0.015\pm 0.016$. The constraint
$\Omega_k=-0.010\pm 0.009$ is obtained by combining the CMB data
with the red luminous galaxy clustering data, which has its own
signature of baryon acoustic oscillations \cite{Eise}. The WMAP
three-year data can (jointly) constrain
$\Omega_k=-0.024_{-0.013}^{+0.016}$ even when allowing  dark
energy with arbitrary (constant) equation of state $w$
\cite{WMAP3}. (The corresponding joint limit from WMAP three-year
data on the equation of state is also impressive,
$w=-1.062_{-0.079}^{+0.128}$).

Due to the results,  it may be interesting to consider other
inflationary universe models where  the spatial curvature is taken
into account \cite{Uzan:2003nk}. In fact, it is interesting to
check if the flatness in the curvature, as well as in the
spectrum, are indeed reliable and robust predictions of inflation
\cite{Linde:2003hc}.

In the context of an open scenario, it is assumed that the
universe has a lower-than-critical matter density and, therefore,
a negative spatial curvature. Several
authors~\cite{re3,re4,re5,re6}, following previous speculative
ideas~\cite{re1,re2}, have proposed alternative  models, in which
open universes may be realized, and their consequences, such as
density perturbations, have been explored~\cite{UMRS}. The only
available semi-realistic model of open inflation with
$1-\Omega\ll\,1$  is rather unpleasant  since it requires a
fine-tuned potential of very peculiar shape~\cite{re6,delherr}.
The possibility to create an open universe from the perspective of
the brane-world scenarios also has been considered~\cite{MBPGD}.

The possibility to have inflationary universe models with
$\Omega>$1 has   been analyzed in \cite{Linde:2003hc,White,Ellis}.
In this paper we would like to describe this kind  of models.

One normally considers the inflation phase driven by the potential
or vacuum energy of a scalar field, whose dynamics is determined
by the Klein-Gordon action. However, more recently and motivated
by string theory, other non-standard scalar field actions have
been used in cosmology. In this context the deep interplay between
small-scale non-perturbative string theory and large-scale
brane-world scenarios has raised the interest in a tachyon field
as an inflationary mechanism, especially in the Dirac-Born-Infeld
action formulation as a description of the D-brane action
\cite{Sen:1998sm}. In this scheme, rolling tachyon matter is
associated with unstable D-branes. The decay of these D-branes
produces a pressureless gas with finite energy density that
resembles classical dust. Cosmological implications of this
rolling tachyon were first studied by Gibbons
\cite{Gibbons:2002md} and in this context it is quite natural to
consider scenarios where inflation is driven by the rolling
tachyon. In  recent years the possibility of an inflationary phase
described by the potential of a tachyon field has been considered
in a quite  diverse topics \cite{Fairbairn:2002yp,Choud}. In the
context of an open inflationary scenario, a universe dominated by
tachyon matter is studied in Ref.~\cite{Balart:2007je}.

In this paper we adopt  the point of view considered by Linde
\cite{Linde:2003hc}  but in which a  tachyon field theory is
considered. More precisely,  we suppose that a closed universe
appears  from nothing at the point in which $\dot{a} = 0$,
$\dot{\phi}=0$, and a potential energy density is $V(\phi)$. We
solve the Friedmann and tachyon field equations considering that
the acceleration of the universe is sufficient for producing
inflationary period. It should be clear from the beginning that
the tachyon potential considered by us  satisfies $dV/d\phi< 0$
for  $\phi > \phi_0$ and $V(\phi \rightarrow \infty) \rightarrow
0$. On the other hand, we assume that the potential becomes
extremely large in the vicinity of $\phi<\phi_0$  since  the
closed universe appeared at this point.

The paper is organized as follows. In Sec.\ref{Sec2} we review
briefly the cosmological equations in the tachyon models.
Sec.\ref{Sec3} presents a toy model in some detail. We get the
value of the tachyon field, when  inflation begins. We also obtain
the probability of the creation of a close universe from nothing.
In Sec.\ref{Sec4} we consider a model with  a tachyonic
exponential potential. In Sec.\ref{Sec5} the cosmological
perturbations are investigated. Finally, in Sec.\ref{Sec6}, we
summarize our results.

\section{Cosmological Equations in  the Tachyon Models}
\label{Sec2}

The action for our model is given by \cite{Sen:2002nu}
 \[ S = S_{grav} + S_{tach}\]
\be = \int \sqrt{-g}\, d^4 x \left[\frac{R}{2
 \kappa}- V(\phi)\sqrt{1 - \partial^\mu \phi \, \partial_\mu \phi} \,
 \right], \label{b-i} \en where $\kappa = 8 \pi G=8 \pi/M_p^2$
 (here $M_p$ represent the Planck mass) and $V(\phi)$ is
 the scalar tachyon potential.

The energy density $\rho$ and pressure $p$ for tachyonic field are
given by  \be \rho =
\frac{V(\phi)}{\sqrt{1-\dot{\phi}^2}}\label{den}, \en and  \be p =
-V(\phi) \sqrt{1-\dot{\phi}^2}\label{pre}, \en respectively.

 The Friedmann-Robertson-Walker metric is described by \be
ds^2=dt^2-a(t)^2\;d\Omega_k^2,\label{met} \en
 where $a(t)$ is the scale
factor, $t$ represents the cosmic time and $d\Omega_k^2$ is the
spatial line element corresponding to the hypersurfaces of
homogeneity, which could be represented as  a three-sphere, a
three-plane or a three-hyperboloid, with values $k=1, 0,-1$,
respectively. From now on we will restrict ourselves to the case
$k=1$ only. Using the metric (\ref{met}) in  the  action
(\ref{b-i}), we obtain the following field equations:  \be
\left(\frac{\dot{a}}{a}\right)^2=-\frac{1}{a^2}+\frac{\kappa}{3}\frac{V(\phi)}{\sqrt{1-\dot{\phi}^2}}\;,
\en \be \frac{\ddot{a}}{a} = -\frac{\kappa}{6}(\rho + 3 p)=
\frac{\kappa}{3}\,\,\frac{V(\phi)}{\sqrt{1-\dot{\phi}^2}} \left(
1-\frac{3}{2}\,\dot{\phi}^2 \right)\,\,\label{ec2}, \en and
 \be \ds
\frac{\ddot{\phi}}{1-\dot{\phi}^2}\,=\,-3\,\frac{\dot{a}}{a}\,\dot{\phi}\,-\,
\frac{1}{V(\phi)}\, \frac{d V(\phi)}{d\phi}\,\,\label{ec3}, \en
where the dot over $\phi$ and $a$ denotes derivative with respect
to the time $t$. For convenience we will use the units in which
$c=\hbar=1$.

\section{Constant Potential}
\label{Sec3}

In the spirit of Ref.~\cite{Linde:2003hc}, we study a closed
inflationary universe, where inflation is driven by a tachyon
field. First let us consider a simple tachyon model with the
following step-like effective potential: $V(\phi)=V = constant$
for $\phi > \phi_0$, and $V(\phi)$ is extremely steep for $\phi
<\phi_0$. We consider that the birth of the inflating closed
universe can be created "from nothing", in a state where the
tachyon field takes the value $\phi_{in} \leq \phi_0$ at the point
with $\dot{a}=0$, $\dot{\phi}=0$  and the potential energy density
in this point is $V(\phi_{in})\geq V=const$ . If the effective
potential for $\phi < \phi_0$ grows very sharply, then the tachyon
field instantly falls down to the value $\phi_0$, with potential
energy $V(\phi_0)= V$, and its initial potential energy
$V(\phi_{in})$ becomes converted to the kinetic energy. Since this
process happens instantly we can consider $\dot{a}=0$, so  that
tachyon field arrives  to the the plateau with a velocity given
by:

 \be \ds
\dot{\phi}_0\,=\, +\sqrt{1-\Big(\frac{V}{V(\phi_{in})}\Big)^2}
\,\,\label{ec2b}. \en

Thus, in order to study the inflation in this scenario, we have to
solve Eqs.~(\ref{ec2}) and (\ref{ec3}) in the interval $\phi \geq
\phi_0$, with initial conditions $\dot{\phi}=\dot{\phi}_0$,
$a=a_0$ and $\dot{a}=0$. These equations have different solutions,
depending on the value of $\dot{\phi}_0$. In particular, if we
insert Eq.~(\ref{ec2b}) into   Eq.(\ref{ec2}), we obtain:

\be \frac{\ddot{a}}{a} = \frac{\kappa}{6}\,V(\phi_{in})
\left[3\left(\frac{V}{V(\phi_{in})}\right)^2 - 1
\right]\,\,\label{ec4}. \en

Then, we  notice that there are three different scenarios,
depending on the particular value of $V(\phi_{in})$. First, in the
particular case when \be \ds \frac{V}{V(\phi_{in})}\,=\,
\frac{1}{\sqrt{3}} \hspace{1cm}  \mbox{or} \hspace{1cm}
\dot{\phi}^2_0 = \frac{2}{3}\label{sta},\en we see that the
acceleration of the scale factor is $\ddot{a}=0$. Since initially
$\dot{a}=0$, then the universe remain static and the tachyon field
moves with constant speed given by Eq.~(\ref{ec2b}).

In the second case we have:

\be \ds 0 \,<\, \frac{V}{V(\phi_{in})}\,<\, \frac{1}{\sqrt{3}}
\hspace{1cm} \mbox{or} \hspace{1cm} \frac{2}{3}
\,<\,\dot{\phi}^2_0 < 1\label{col}.\en In this case the universe
start  moving with negative acceleration  ($\ddot{a}<0$) from the
state $\dot{a}=0$. Then, in the tachyon field equation describing
negative friction, we have a   term  which make the moving of
$\phi$ even  faster, so that $\ddot{a}$ becomes more negative.
This universe  rapidly collapses.

The third case corresponds to:

\be \ds \frac{1}{\sqrt{3}}\,<\, \frac{V}{V(\phi_{in})}\,<\,1
 \hspace{1cm} \mbox{or} \hspace{1cm}0\,<\,\dot{\phi}^2_0 <
\frac{2}{3}\label{co2}.\en

In this case we have $\ddot{a}>0$, and   the universe enters into
an inflationary stage.

In what follows, we are going to make a simple analysis of the
cosmological equations of motion for cases where the condition
(\ref{co2}) is satisfied. The tachyon field satisfies the equation

\be \ds \frac{\ddot{\phi}}{1-\dot{\phi}^2} +
3\,\frac{\dot{a}}{a}\,\dot{\phi} = 0\,\,\label{Vconst}, \en which
implies  \be \ds \dot{\phi}^2(t) = \frac{1}{1 + C
a^6(t)}\,\,\label{phidot}, \en where $C$ is a positive integration
constant defined as  \be \ds C = \frac{1 - \dot{
\phi}^2_0}{\dot{\phi}^2_0 a^6_0}\,, \en and with $\dot{\phi}^2<1$.

Here $\dot{\phi}_0$ is the initial velocity of the field $\phi$,
immediately after it rolls down to the flat part of the potential.
The effect of the tachyonic field in this model is reflected in
the change of the slope of the tachyonic field $\phi$, when
compared  to standard case, where $\dot{\phi}=\dot{\phi}_0
[a_0/a(t)]^3$.

The behavior of the tachyon field expressed by Eq.(\ref{phidot})
implies that the evolution of the universe rapidly falls into an
exponential regimen (inflationary stage) where the scale factor
becomes $a\sim e^{Ht}$ with $H=\sqrt{\frac{\kappa V}{3}}$. When
the universe enters the inflationary regime, the tachyon field
moves by an amount  $\Delta \phi_{inf}$ and then stops. From
Eq.~(\ref{phidot}) we get:

\[\ds \Delta \phi_{inf} = \frac{1}{3
H}\sinh^{-1}\left(\frac{1}{\sqrt{C}}\right)\]
\be = \frac{1}{3 H}
\ln \left(\frac{1}{\sqrt{C}} + \sqrt{1 +
\frac{1}{C}}\right)\,\,\label{fi-inf}. \en

Note that, when $\dot{\phi}_0 \ll 1$  ($a_0= 1$), we obtain
\mbox{$\Delta \phi_{inf}\approx \frac{\dot{\phi}_0}{3 H}$}, which
coincides with the result obtained in \cite{Linde:2003hc}.

At early time, before inflation take place, we can write
conveniently the equation for the scale factor as follows: \be
\ddot{a}(t) = \frac{2\kappa}{3}\, V\, a(t)\,
\beta(t)\,\,\label{beta1}. \en

 Here, we have introduced a small
 time-dependent dimensionless  parameter $\beta (t)$: \be \beta(t) =
 \frac{1}{2}\, \frac{1}{\sqrt{1-\dot{\phi}^2}} \left( 1 - \frac{3}{2}
 \dot{\phi}^2\right)\label{def-bet}. \en
Certainly $\beta(t)\ll 1$ when $\dot{\phi}^2(t)\rightarrow 2/3$.

Now we proceed to make a  simple  analysis of the scale factor
equation (\ref{ec2}) and the tachyon Eq.(\ref{ec3}) for
$\beta(0)\equiv\beta_0 \ll 1$.

At the beginning of the process, we have $a(t) \approx a_0$ and
$\beta(t) \approx \beta_0$, then Eq.(\ref{beta1}) takes the form:
\be \ddot{a}(t) = \frac{2\kappa}{3}\, a_0 V
\beta_0\,\,\label{beta}, \en and hence for small $t$ the solution
of this equation is given by
 \be \ds
a(t) = a_0 \left(1 + \frac{\kappa \beta_0  V}{3} t ^2\right).
\label{adet} \en

From Eqs. (\ref{phidot}) and (\ref{adet}) we find that at a time
interval where $\beta(t)$ becomes twice as large as $\beta_0$,
$\Delta t_1$ is given by

\[\Delta t_1=\left[\frac{(1-\dot{\phi}_0^2)^3}{2\kappa\,
V\dot{\phi}_0^2}\right]^{1/2} \times \] \be
\left[(2-3\dot{\phi}_0^2)+
\frac{\dot{\phi}_0^2}{2}\;(11-6\dot{\phi}_0^2)+
\frac{3}{2}\dot{\phi}_0^4\right]^{-1/2}.\label{tiempo} \en

Consequently the tachyonic field increases  by the amount \be \ds
\Delta \phi_1 \sim  \dot{\phi}_0 \, \Delta t_1 \sim
\frac{1}{\sqrt{\kappa\,V}}\,\,\label{bexp}, \en  where we have
kept only the first term in the expansion of $\Delta t_1$. Note
that this result depends on the value of $V$, i.e. the increase of
the tachyonic field  is less restrictive than the one used in the
standard scalar field in which $\Delta \phi_1 =const.\sim
-1/(2\sqrt{3\pi})$ \cite{Linde:2003hc}. After the time $\Delta t_2
\approx \Delta t_1$, where now the tachyonic field increases by
the amount $\Delta\phi_2 \approx \Delta\phi_1$, the rate of growth
of $a(t)$ also increases. This process finishes when $\beta(t)
\rightarrow 1/2$. Since at each interval $\Delta t_i$ the value of
$\beta$ doubles, the number of intervals $n_{int}$ after which
$\beta(t) \rightarrow 1/2$ is
 \be \ds
n_{int} = - 1 - \frac{\ln \beta_0}{\ln 2} \,\,\label{n}. \en
Therefore, if we know the initial velocity of the tachyon, we can
estimate the value of the tachyon field at which the inflation
begins:

\be \ds \label{ficero} \phi_{inf} \sim \, \phi_0  - \!\left(1 +
\frac{\ln \beta_0}{\ln 2} \right) \!\frac{1}{\sqrt{\kappa V}}
\,\,. \en  This expression indicates that our result for
$\phi_{inf}$ is sensitive to the choice of particular value of the
potential energy $V$, apart from the initial velocity of the
tachyonic field $\phi$ immediately after it rolls down to the
plateau of the potential energy.

Note that if the tachyon field starts its motion with a small
velocity, the inflation begins immediately. However, if the
tachyon moves  with a large initial velocity the inflation is
delayed, but once the inflation begins, it never stops. This can
be explained by the constancy of the potential, and  as we will
see,in the next section this particular problem disappears when we
consider a more realistic tachyon model.

We return to  describing a model of quantum creation for a closed
inflationary universe model. The probability of the creation of a
closed universe from nothing  is given by Ref.\cite{Koya}

\be P\sim e^{-2\mid S
\mid}=\exp\left(\frac{-\pi}{H^2}\right)\sim\exp\left(\frac{-3
\pi}{\kappa\;V(\phi)}\right). \en

 We first estimate the conditional probability that the universe
is created with an energy density equal to
$\sqrt{3}\;V-\beta_0\,V$. Assuming  that this energy  is smaller
than $V(\phi_{in})=\sqrt{3}\;V$,  for the probability we get

\[\ds P \sim e^{-2\mid S \mid} \sim \exp\left(-
\frac{3M_p^4}{8(\sqrt{3}-\beta_0)V} +
\frac{3M_p^4}{8\sqrt{3}V}\right)\] \be \sim \exp
\left(-\frac{M_p^4 \beta_0}{8 V}\right) \,\,\label{prob}, \en
where we have used that $\dot{\phi}^2\ll 1$. This implies  that
the process of quantum creation of an inflationary universe is not
exponentially suppressed if $\beta_0 < 8V/M_p^4$.

\section{Exponential potential}
\label{Sec4}

\begin{figure}
\hspace{1cm}\resizebox{0.35\textwidth}{!}{%
  \includegraphics{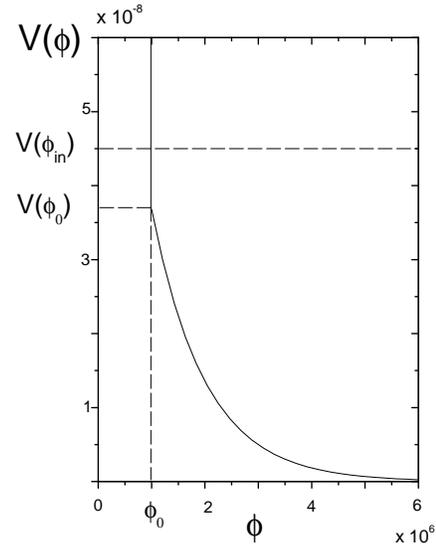}
}
\caption{The plot shows the tachyonic potential as a function of
the tachyonic field $\phi$. We have taken $V_0=10^{-7}
\kappa^{-2}$ and $\lambda=10^{-5}\kappa^{-1/2}$ in  units where
$\kappa$=1.}
\label{pot1}     
\end{figure}

Now we proceed with a more realistic case,  a  model in which the
effective potential is  given by

\be \label{pot} V(\phi) \simeq V_0\,e^{-\lambda \phi}, \en where
$\lambda$ and $V_0$ are free parameters and the parameter
$\lambda$ is related with the tachyon mass
\cite{Fairbairn:2002yp}, in the following we will take $\lambda
> 0$ (in units $M_p$).   We will also assume that the effective potential
sharply rises to indefinitely large values in a small vicinity of
$\phi=\phi_0$ , see Fig.\ref{pot1}.


We assume that the whole process is divided in three parts. The
first part  corresponds to the creation of the (closed) universe
``from nothing" in a state where the tachyon field takes the value
$\phi_{in} \leq \phi_0$ at the point with $\dot{a}=0$,
$\dot{\phi}=0$,  and where the potential energy is $V(\phi_{in})$.
If the effective potential for $\phi\ < \phi_0$ grows very
sharply, then the tachyon field instantly falls down to the value
$\phi_0$, with potential energy $V(\phi_0)$, and the initial
potential energy becomes converted to the kinetic energy, see the
previous section. Then we have:

\be \dot{\phi}^2_0= 1-
\left(\frac{V(\phi_0)}{V(\phi_{in})}\right)^2. \en

Following the discussion of the previous section we suppose that
the following initial condition is satisfied:

\be \label{cond} V(\phi_0) < V(\phi_{in}) < \sqrt{3}\,V(\phi_0),
\en which guarantees that the model arrives  to an inflationary
regimen. As it was mentioned previously, in all other cases the
universe remains either static, or it  collapses rapidly.

The second and third parts of the process  are described by
Eqs.~(\ref{ec2}) and (\ref{ec3}) in the interval $\phi \geq
\phi_0$ with initial conditions $\dot{\phi}=\dot{\phi}_0$, $a=a_0$
and $\dot{a}=0$. In particular, the second part of the process
corresponds to the motion of the tachyon field before the
beginning of the  inflation stage, and it is well described by the
following approximation of the equations of motion:

\begin{eqnarray}
\frac{\ddot{\phi}}{1-\dot{\phi}^2} =
-3\,\frac{\dot{a}}{a}\,\dot{\phi}\,,
\label{t1}\\
\nonumber \\
\label{t2} \ddot{a} = \frac{2\kappa}{3}\,a\,V(\phi)\,\beta(t),
\end{eqnarray}
which  $\beta(t)$ satisfying  $\beta(t)\ll 1$, as before.

The third part corresponds to the stage of inflation where
$\dot{\phi}$ is small enough and the scale factor $a(t)$ grows  up
exponentially. This part is well described by the following
approximation of the equation of motion \cite{Sami:2002fs}:

\be \ds 3\,\frac{\dot{a}}{a}\,\dot{\phi} = -\frac{1}{V}\,
\frac{dV}{d\phi}\,\,\label{inf1}, \en
\begin{equation}
\label{inf2} \ddot{a} = \frac{\kappa}{3}\,a\,V(\phi)\,.
\end{equation}

In summary, the whole process could be described as follows: the
tachyon field starts its motion   at $\phi = \phi_{in}$, with
$\phi_{in}<\phi_0$, then the field immediately moves to the value
$\phi_0$ and acquires a non-null velocity $\dot{\phi}_0$. After
that the tachyon field starts to move subjected to the  equations
of motion, and  the potential  in Eq.~(\ref{ec3}) can be neglected
so that  the only contribution to the evolution of $\phi$ is the
dissipative term.  Therefore, during the period when the condition
(\ref{cond}) is satisfied, $\phi$ satisfies Eq.(\ref{phidot}),
which means that $\dot{\phi}$ drops down while the size of the
universe is increasing.  This initial behavior for $\dot{\phi}$ is
in very good agreement with the phase portrait (with numerical
results) for tachyonic cosmology described in Ref.\cite{Zong}. As
a result of this process  we arrive to an inflationary regimen,
described by the Eqs. (\ref{inf1}) and (\ref{inf2}).

Now, we are going  to describe the process in  more details. Let
us consider  the second stage, where the tachyon field satisfies
Eq.(\ref{t1}) and the scale factor satisfies Eq.(\ref{t2}).
Following the scheme of section \ref{Sec3} we solve the equation
for $a(t)$ by considering $\beta(t) \ll 1$. Then at the beginning
of the process, when $a\approx a_0$ and $\beta \approx \beta_0$,
Eq.~(\ref{t2}) takes the form:

\be \ddot{a}(t) = \frac{2\kappa}{3}\, a_0 V(\phi_0)
\beta_0\,\,\label{beta}, \en and the tachyon field satisfies
Eq.(\ref{phidot}). The amount of the increasing of the tachyon
field during the time $\Delta t$ which make the value of $\beta$
two times grater than $\beta_0$ is:

\be \Delta \phi \approx \frac{1}{\sqrt{\kappa V(\phi_0)}}. \en

This process continues until $\dot{\phi}$ is small enough so that
the universe begins to expand  in an exponential way,
characterizing the inflationary era. we take that the  inflation
begins when $\beta(t)$ approaches to $1/2$.  Then, according to
our previous result,  the tachyon field gets the value:

\be \ds \label{inf3} \phi_{inf} \sim \, \phi_0  - \!\left(1 +
\frac{\ln \beta_0}{\ln 2} \right) \!\frac{1}{\sqrt{\kappa
V(\phi_0)}} \,\,. \en

In order to find analytical solution to the equation of the
tachyon field in the inflationary era, we are going to focus to
the approximation of flat space for the Friedmann equations. Then
we can use the result of Ref.\cite{Sami:2002fs}.

The tachyon field satisfies the following equation:

\be \label{sami1} \dot{\phi}=\frac{\lambda}{3\gamma}\,e^{\lambda
\phi/2}, \en where $\gamma^2 = V_0\kappa/3$. Notice that
$\dot{\phi}$ increases during the inflationary era, what differs
from $\dot{\phi}$  in the previous period, see Eq.(\ref{phidot}).
The scale factor has the following behavior during this period:

\be \label{sami2} \frac{a(t)}{a_i} = e^{\gamma
t[C-(\lambda^2/12\gamma)t]}, \en where $a_i$ is the value of the
scale factor at the beginning of inflation and
$C=e^{-\frac{\lambda \phi_{inf}}{2}}$. From Eq.(\ref{sami2}) we
notice that the scale factor passes through an inflection point
which marks the end of inflation. This happen when
$\dot{\phi}_{end} = \sqrt{2/3}$, which implies  that the value of
the potential at the end of inflation is
$V_{end}=\frac{\lambda^2}{2\kappa}$. The values of the tachyon
potential at the beginning and at the end of inflation are related
by the number of the e-folds ${\cal N}$, see
Ref.\cite{Sami:2002fs}:

\be \label{sami3} (2 {\cal N} +1)V_{end} = V(\phi_{inf}). \en

Then, by using Eq.(\ref{inf3}) and Eq.(\ref{sami3}), we can
related the value of $\beta_0$ with the number of e-folds, ${\cal
N}$

\be \label{betaN} \beta_0 = \frac{1}{2}\left[\frac{(2 {\cal N}
+1)}{V(\phi_0)}\,\frac{\lambda^2}{2\kappa}
\right]^{\frac{\sqrt{\kappa V(\phi_0)}}{\lambda}\ln(2)}. \en

Note that, just as before, if the tachyon field starts its motion
with sufficiently small velocity (large $\beta$), inflation begins
immediately and we have a large number of e-folds. On the other
hand, if the tachyon field  it starts  with a large initial
velocity $\dot{\phi}_0 \sim \sqrt{2/3}$,  corresponding  to
$\beta_0 \approx 0$, the beginning of inflation takes more time
and we obtain a lower value of ${\cal N}$. Eventually, if
$\beta_0$ is too small, we can arrive to the situation where
\begin{equation}
\label{noinflation} V(\phi_{inf})< V_{end} =
\frac{\lambda^2}{2\kappa}\,,
\end{equation}
and  the universe can never inflate.

As an example, we  take a particular set of  the parameters
appearing  in the tachyon potential (\ref{pot}). We also use the
COBE normalized value for the amplitude of scalar density
perturbations in order to evaluate $\lambda$
 \cite{Sami:2002fs}, thus we have $\lambda = 10^{-5}\kappa^{-1/2}$ and
we  take $V_0 = 10^{-7} \kappa^{-2}$.

\begin{figure}
\hspace{1 cm}\resizebox{0.35\textwidth}{!}{%
  \includegraphics{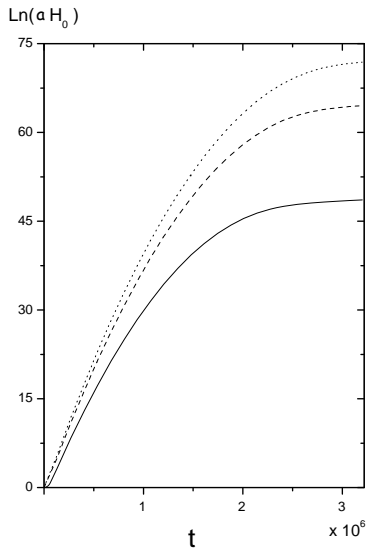}
}
\caption{This plot shows the number of e-folds as function of
cosmological time t, for different initial values of
$\dot{\phi}_0^2$. We  have taken   $\kappa$=1.}
\label{CombN1}       
\end{figure}

Note that, if the field starts with a large velocity
$\dot{\phi}_0$, the universe starts to inflate  in a late time and
a lower value of the number of  e-folding is obtained (see
Fig.(\ref{CombN1})).

Now, let us analyze  the quantum probability of creation of
considered sort of universe. From the discussion of section
\ref{Sec3} we know that the probability of the universe with
$\beta_0 \neq 0$ will be exponentially  suppressed, unless the
universe is created very close to the threshold value
$V(\phi_{in})=\sqrt{3}V(\phi_0)$, with

\be \label{cond2} \beta_0 < \frac{\kappa^2 \,V(\phi_0)}{8\pi^2}.
\en

If we assume that $\phi_0 \sim 10^5 \kappa^{1/2}$,  then  in order
to satisfy Eq.(\ref{cond2}) we have $\beta_0 < 2.2 \cdot
10^{-10}$. Following Ref.\cite{Linde:2003hc} we can argue that the
probability for start with the value $\beta_0 \ll 2.2 \cdot
10^{-10}$ is suppressed,  due to the small phase space
corresponding to these values of $\beta_0$. Thus, it is most
probable to have $\beta_0 \sim 2.2 \cdot 10^{-10}$, and  in that
case if we set  $\phi_0 = 1.1 \cdot 10^{5} \kappa^{1/2}$ which
satisfies the condition Eq.(\ref{cond2}) and we obtain ${\cal N} =
60$,  this leads to $\Omega=1.1$. On the other hand, if we take
$\phi_0 = 0.5 \cdot 10^5 \kappa^{1/2}$, we get ${\cal N} = 171$
and the universe becomes flat.

\section{Perturbations }
\label{Sec5}
 Even though the study of scalar density perturbations
in closed universes is quite complicated, it is interesting to
give an estimation of the standard quantum scalar field
fluctuations in this scenario. In particular, the spectra of
scalar perturbations for a flat space, generated during tachyon
inflation, expressed in terms of the slow-roll parameters defined
in Ref. \cite{Schwarz:2001vv}, becomes \cite{ste-ver}:
\be \frac{\delta\,\rho}{\rho}\, = \left[1 - 0.11\, \epsilon_1 +
0.36 \, \epsilon_2 \right]\frac{\kappa\,H}{2\pi\,\sqrt{2
\epsilon_1}}\,, \label{ec10}
 \en
where the slow-roll parameters are given by:
\begin{eqnarray}
 \epsilon_1 &\simeq&\frac{1}{2\,\kappa}\frac{(V\!,_{\phi})^2}{V^3}\,, \\
\epsilon_2 &\simeq& \kappa^{-1}\,\Big[-2\frac{V,_{\phi\phi}}{V^2}+
3 \frac{(V\!,_{\phi})^2}{V^3}\Big].\label{e2}
 \end{eqnarray}

Certainly, in our case, Eq.(\ref{ec10}) is an approximation and
must be supplemented by several different contributions in the
context of a closed inflationary universe \cite{Linde:2003hc}.
However, one may expect that the flat-space expression gives a
correct result for $N>3$.

If one interprets perturbations produced immediately after the
creation of closed universe (at $N\sim O(1)$) as perturbations on
the horizon scale $l\sim 10^{28} cm$, then the maximum at $N\sim
10$ would correspond to the scale $l\sim 10^{24} cm$, and the
maximum at $N\sim 15$ would correspond to the scale $l\sim 10^{22}
cm$, which is similar to the galaxy scale.

One interesting parameter to consider is the so-called spectral
index $n_s$, which is related to the power spectrum of density
perturbations $P^{1/2}_{\cal R}(k)$. For modes with a wavelength
much larger than the horizon ($k \ll a H$), the spectral index
$n_s$ is an exact power law, expressed by $P^{1/2}_{\cal R}(k)
\propto k^{n_s-1}$, where $k$ is the comoving wave number. It is
also interesting to give an estimate of the tensor spectral index
$n_T$.
In tachyon inflationary models the scalar spectral index and the
tensor spectral index are given by
\begin{equation}
n_s=1-2\epsilon_1 - \epsilon_2\label{ns},
\end{equation}
and $n_T = -2\epsilon_1$, in the slow-roll approximation
\cite{ste-ver}.

One of the  features of the 3-year data set from WMAP is that it
hints at a significant running in the scalar spectral index
$dn_s/d\ln k=\alpha_s$ \cite{WMAP3}. From Eq.(\ref{ns}) we obtain
that the running of the scalar spectral index for our model
becomes

\begin{equation}
\alpha_s=\frac{d n_s}{d\ln
k}\simeq\;2\frac{V\;\epsilon_{1}}{V_{,\;\phi}}
[2\;\epsilon_{1\;,\;\phi}+\epsilon_{2\;,\;\phi}],\label{dnsdk}
\end{equation}
where we have used that $d\ln k=-dN$. Using the exponential
potential from Eq.(\ref{dnsdk}) we find that,
\begin{equation}
\alpha_s=\frac{d n_s}{d\ln
k}\simeq\;-2\frac{\lambda^4}{\kappa^2\;V_0^2}\;e^{2\lambda\;\phi}=
-2\frac{\lambda^4}{\kappa^2\;V(\phi)^2}\;.\label{as2}
\end{equation}

Note  the difference that occurs  with respect to a standard
scalar field (with an exponential potential) where $\alpha_s=0$,
since $n_s=Cte.=1-M_P^2\lambda^2$ \cite{pp1}.

In models with only scalar fluctuations, the marginalized value
for the derivative of the spectral index is approximated to
$dn_s/d\ln k=\alpha_s \sim -0.05$ for WMAP three-year data only
\cite{WMAP3}.

 Noted that, from Eq. (\ref{as2}), the scalar potential becomes
$V(\phi_{*})\sim 6.3 \lambda^2/\kappa$, where $\phi_{*}$
represents the value of the tachyon field when the scale
$k_0=0.002$ Mpc$^{-1}$  leaves  the horizon. For $\lambda\sim
10^{-5}\kappa^{-1/2}$, we have for the scalar potential, when the
scale is $k_0$ was leaving the horizon, becomes, $V(\phi_{*})\sim
10^{-9} \kappa^{-2}$. This value of the scalar potential is in
agreement with Ref.\cite{Bhatt} where a chaotic potential  with a
standard scalar field is used.

Using the WMAP three-year data\cite{WMAP3} and  the SDSS large
scale structure surveys \cite{Teg},  an upper bound
$\alpha_s(k_0$) has been found, where $k_0$=0.002 Mpc$^{-1}$
corresponds to $L=\tau_0 k_0\simeq 30$, with the distance to the
decoupling surface $\tau_0$= 14,400 Mpc. SDSS measures galaxy
distributions at red-shifts $a\sim 0.1$ and probes $k$ in the
range 0.016 $h$ Mpc$^{-1}$$<k<$0.011 $h$ Mpc$^{-1}$. The  recent
WMAP three-year data results give the values for  the scalar
curvature spectrum $P_{\cal
R}(k_0)\equiv\,25\delta_H^2(k_0)/4\simeq 2.3\times\,10^{-9}$ and
the spectral index $n_s\simeq 0.95$. These values allow  us to
find the constraints  on the  parameters of our model.
Furthermore,  from the numerical solution we can obtain their
values. In particular, for  $k_0$=0.002 Mpc$^{-1}$, we have $n_s
\approx 0.96$ and $n_T \approx -0.03$. Notice that those indices
are very close to the Harrison-Zel'dovich spectrum \cite{Ha-Ze}.

\section{Conclusion and Final Remarks}
\label{Sec6}

In this work we have studied  a closed inflationary universe model
in a  tachyonic field theory. In the context of Einstein's GR
theory, this model was studied by Linde \cite{Linde:2003hc}.
Firstly, we have assumed  a tachyon scalar potential to be
constant. Secondly,  we have analyzed  a closed universe model
with a tachyon exponential potential. For these two parts, in
which the potential is very sharp at small values of the field,
$\phi<\phi_0$, we have found extra ingredients in the tachyonic
theory,  compared to its analog in standard scalar field theory.
Specifically, we obtain a large stage of initial inflation for a
closed universe, and this stage  depends on   the value of the
potential energy $V$. In this way, we have found that our model,
which takes into account a tachyonic theory, is less restrictive
that the one used in standard scalar field theory.

Also, we have found that, after  the tachyon field starts  moving
subjected to the equations of motion, the potential term in
Eq.~(\ref{ec3}) becomes irrelevant, and the only contribution to
the evolution of $\phi$ is the dissipative term. Therefore, during
this period, $\phi$ satisfies the Eq.(\ref{phidot}), which means
that $\dot{\phi}$ drops down, since the size of the universe is
increasing in this period, if the condition (\ref{cond}) is
satisfied. This initial behavior for $\dot{\phi}$ is in good
agreement with the phase portrait (with numerical results) for the
tachyonic cosmology  described in Ref.\cite{Zong}.

We have also found that, for an exponential
potential, the inclusion of the tachyonic field changes some
characteristic of the running spectral index $\alpha_s$ and
becomes $\alpha_s\neq 0$, in contrast to the standard case where
$\alpha_s=0$. From the normalization of the WMAP three-year data,
the potential becomes of the order of $V(\phi_{*})\sim
10^{-10}M_p^4$ when it leaves the horizon at the scale of
$k_0=0.002$ Mpc$^{-1}$. In particular, we expect that the Planck
mission will significantly enhance our understanding of $\alpha_s$
by providing high quality measurements of the fundamental power
spectrum over a large wavelength range  that WMAP.
Summarizing, we have been successful in describing a closed inflationary universe
in a  tachyon field  theory.

\section*{Acknowledgments}

We thank  Olivera Miskovic  for a careful reading of the
manuscript. L. B. is supported from PUCV through Proyecto de
Investigadores J\'ovenes a{\~n}o 2006. S. del C. is supported by
the COMISION NACIONAL DE CIENCIAS Y TECNOLOGIA through FONDECYT
Grant N$^{0}$. 1070306, and also was partially supported by PUCV
Grant N$^0$. 123.787/2007. R. H. is supported by the ``Programa
Bicentenario de Ciencia y Tecnolog\'{\i}a" through the Grant
``Inserci\'on de Investigadores Postdoctorales en la Academia"
\mbox {N$^0$. PSD/06}. P. L. is supported by the COMISION NACIONAL
DE CIENCIAS Y TECNOLOGIA through FONDECYT Postdoctoral Grant
N$^0$. 3060114.

%
%

\end{document}